\documentclass[preprint,12pt,authoryear]{elsarticle}

\usepackage{amssymb}

\newcommand{\araa}{Annu. Rev. Astron. Astrophys.}   
\newcommand{\apj}{Astrophys. J.}   
\newcommand{\apjl}{Astrophys. J. Lett.}   
\newcommand{\aap}{Astron. Astrophys.}   
\newcommand{\mnras}{Mon. Not. R. Astron. Soc.}   
\newcommand{\nat}{Nature} 

\journal{High Energy Astrophysics}

\begin{document}

\begin{frontmatter}



\title{\bf An explanation for the radio variation associated with the Vela pulsar glitch occurred on December 12 2016}


\author{Shuang Du}
\ead{dushuang@pku.edu.cn}

\address{School of Mathematics and Computer Science, Tongling University, Tongling, 244000, Anhui, China}

\begin{abstract}
The elaborate observation of the single radio pulses of Vela pulsar around the pulsar glitch that occurred on December 12, 2016 reveals that the physical mechanism associated with this glitch exert a profound influence on the pulsar's magnetosphere.
According to the evolution of these pulses, we propose a scenario regarding how the pulsar magnetic field might undergo alterations within the framework of the inner gap model.
We deduce that the liberation of the free energy within Vela pulsar results in the emergence of new magnetic multipole components.
The progressively developing multipole components cause the magnetic field lines in a section of the polar cap region to become increasingly curved, ultimately resulting in the observed pulse broadening and pulse missing.
At last, we discuss the possible connection between magnetic variations and fast radio bursts according to the inspiration of the presented picture.
\end{abstract}




\end{frontmatter}


\section{Introduction}
\label{introduction}

The stable periodic pulsed emissions emanating from pulsars indicate that, for a rotating pulsar, a stable pulsar magnetosphere comprises both an open-field-line region and a close-field-line region \citep{1969ApJ...157..869G}, and the open-field-line region may maintain
certain gaps within it (\citealt{1975ApJ...196...51R,1981ApJ...248.1099A,1986ApJ...300..500C,
2004ApJ...606L..49Q,2008ApJ...683L..41B}).
This is one of the widely accepted aspects on the stable structure of pulsar magnetic fields.
On the other hand, the long-term timing observations (\citealt{1993MNRAS.265.1003L,2015MNRAS.446..857L}) also shed the information on the evolution of pulsar magnetic fields (see the effects summarized in \citealt{2016ApJ...833..261B,2022Symm...14..130G,2024MNRAS.528.5178S}).
However, extracting meaningful information from these observations is challenging, owing to the complexities involved in the coupling between pulsar magnetic fields and rotational inertias.
Fortunately, some transient X-ray phenomena of pulsars indicate the decay and variation of pulsar magnetic fields (e.g., \citealt{1995MNRAS.275..255T,1996ApJ...473..322T}; see \citealt{2015RPPh...78k6901T,2017ARA&A..55..261K} for reviews),
especially with regards to the toroidal components of these fields  (\citealt{2001ApJ...561..980T,2009ApJ...703.1044B,2020MNRAS.497.2680T}; see also \citealt{1998ApJ...505L.113K}).
Additionally, an elaborate observation in the radio band may provide an opportunity to achieve a deeper understanding of this issue.

\cite{2018Natur.556..219P} found that the pulse profile and polarization of the Vela pulsar exhibited rapid variations in tandem with the glitch event that occurred on December 12, 2016. Specifically, among a consecutive series of single pulses, one pulse exhibited an unusually broad profile, the subsequent pulse was missing, and the two pulses thereafter displayed unexpectedly low linear polarization.
Palfreyman et al. considered that alterations in the magnetic field of Vela pulsar were likely the cause behind the glitch, as well as the changes in pulse characteristics,
since magnetic fields may act as barriers against the migration of superfluid vortices \citep{1999MNRAS.307..365S}.
However, this explanation lacks specific details regarding the variations in pulse shape.

Although, the precise physical mechanism responsible for initiating glitches remains elusive,
the magnetic variation indicated by the change of the linear polarization of single pulses
should be reliable, given that linear polarization typically relies on ordered magnetic fields.
In this manuscript, the magnetic variation indicated by the depolarization of single pulses is treated as the precondition, while the other two clues that the pulse missing and pulse broadening are utilized to investigate how the magnetic variation can be.
In the following section, we present a suggestive physical picture
grounded in the popular framework of pulsar magnetospheres \citep{1969ApJ...157..869G,1975ApJ...196...51R}.

\section{The basic framework}\label{sec2}
As mentioned in Introduction, the depolarization of single pulses serves as the indicator of the magnetic variations of Vela pulsar.
Given its correlation with the glitch, the underlying cause of the magnetic variations could be the alterations of the velocity field in Vela pulsar.
No matter which mechanism triggers the glitch (e.g., \citealt{1975Natur.256...25A,1984ApJ...276..325A}; see \citealt{2015IJMPD..2430008H,2022Univ....8..641Z} for reviews),
there must be a transfer of angular momentum within the pulsar since glitches are sudden spin-up of pulsars. Consequently, it is natural to consider that the velocity field undergoes a change in tandem with the glitch.
Since the magnetic field should be coupled to the velocity field within a pulsar,
any variation in the latter inevitably leads to corresponding change in the former.

The remaining crucial problem is: how variations in the magnetic field give rise to pulse broadening and pulse missing.

\begin{center}
\begin{figure*}
\centering
\includegraphics[scale=.45]{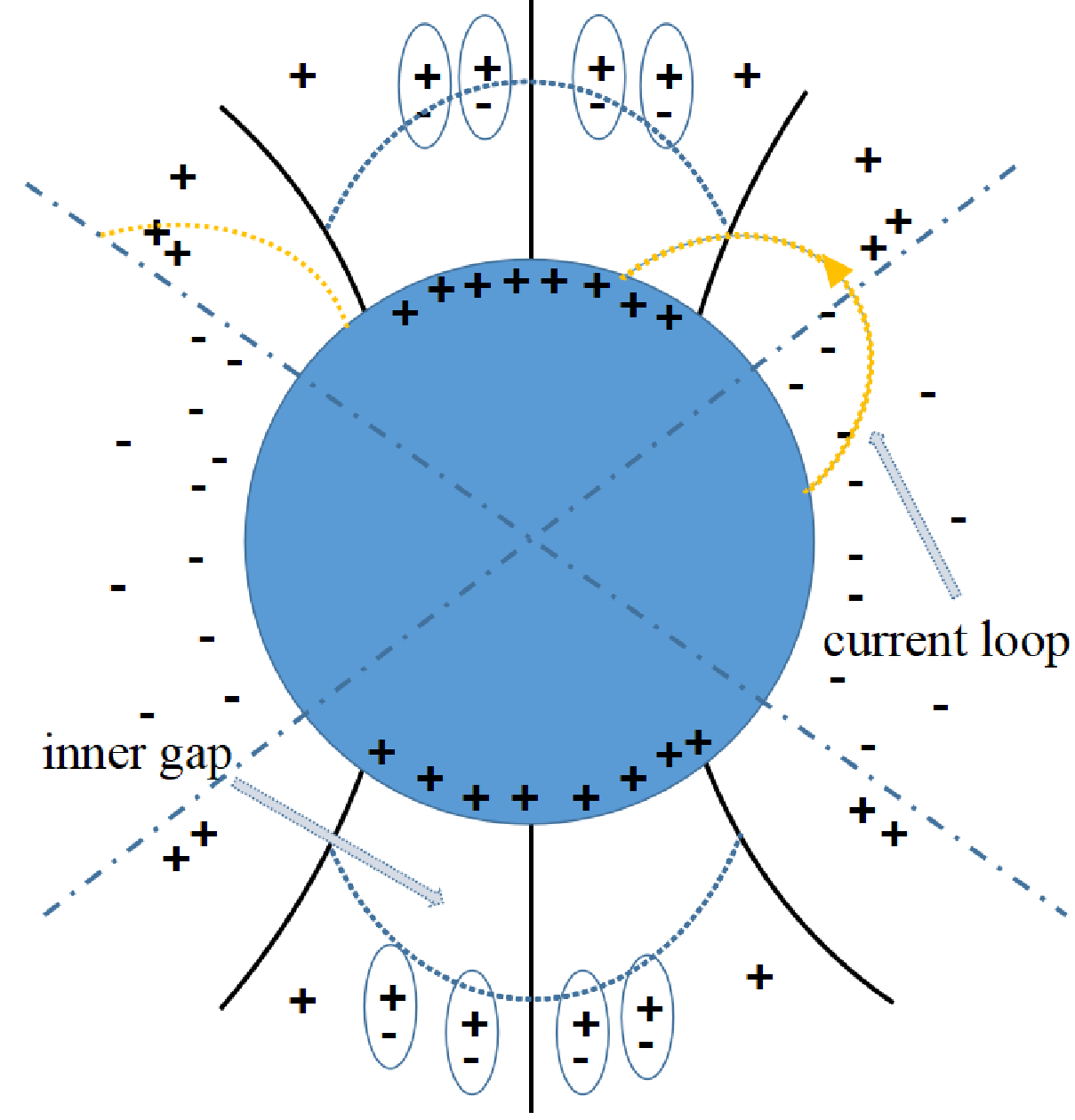}
\caption{
The sketch map of the variation of the magnetic field on Vela pulsar. The blue circle is the pulsar. The dot-dashed lines are the coboundaries of the inner gap.
The positive and negative signs represent the positive and negative charges, respectively. The dashed lines are the null charge surfaces.
The solid lines are magnetic field lines. The dotted lines are the newly generated magnetic field lines due to the release of free energy.
The newly created close field lines cause the movement of charges to form a closed loop, thereby hindering the development of the gap.
}
\label{f1}
\end{figure*}
\end{center}

Based on the inner gap model \citep{1975ApJ...196...51R}, for a successful radio pulse,
the potential across the gap must increase to the critical threshold necessary for the generation of sparks.
The Gaussian-like pulse profile indicates that the region closer to the radio beam's center should be more conducive to sparking; otherwise, a uniform distribution of spark frequency in the gap would result in the pulse profile resembling a square wave.
Figure \ref{f1} illustrates one scenario response for the Gaussian-like pulse profile, in which the maximum gap height, as well as the potential across the gap, undergos a decrement as the distance from the beam's center escalates.
Alternatively, another scenario may be that the critical threshold for sparks increases with the distance from the beam's center.
In both scenarios, an effective strategy to broaden the pulse profile involves lowering the threshold voltage necessary for spark generation.
To accomplish this objective via magnetic variations, multipole field lines are required.
For an inner gap, the critical voltage required to trigger sparks can be estimated as \citep{1975ApJ...196...51R}
\begin{eqnarray}\label{o1}
V_{\rm c}=\frac{2\pi |\mathbf{B}_{\rm s}|}{c}h_{\rm c}^{2}\approx 5\times 10^{9} \rho_{6}^{4/7}P^{-1/7}B_{12}^{-1/7}{\rm V},
\end{eqnarray}
where $\rho$ is the radius of curvature of magnetic field lines and $\rho_{6}=\rho/10^{6}{\rm cm}$,
$\mathbf{B}_{\rm s}$ is the surface magnetic field and $B_{12}=|\mathbf{B}_{\rm s}|/10^{12}{\rm G}$,
$P$ is the rotation period, and
\begin{eqnarray}
h_{\rm c}\approx 5\times 10^{3}\rho_{6}^{2/7}P^{3/7}B_{12}^{-4/7}
\end{eqnarray}
is the critical thickness of the gap for sparks.
As emphasized in \cite{1975ApJ...196...51R}, the typical value that $\rho\sim 10^{6}\rm cm$ is not the case of a pure dipole magnetic field, and higher multipole components are required to contribute most strongly near the pulsar surface (see Appendix A for more discussions).
We note that toroidal components of pulsar magnetic fields have the potential to liberate free energy.
When a portion of this energy is converted into the generation of a new poloidal component, particularly curved multipole components, at the place of the gap where no spark had previously occurred, sparks may be triggered at this place due to the reduction of the critical spark voltage.
Therefore, the region where sparks can occur is broadened, as well as the pulse profile.\footnote{As the multipole component is close to the pulsar surface, the pulsed X-ray emission should not be affected evidently.}

Furthermore, if the newly formed multipole components can be further developed, the field lines of the poloidal component could be sufficiently curved and even changed into close field lines. Under this extreme case, a portion of the open-field-line region may be transformed into a closed-field-line region, thereby the formation of the gap within this altered region is prevented. When the line of sight sweeps over this newly formed close-field-line region, the pulse will be missed.

In the following section, we analytically discuss the possibility of the above picture.

\section{Magnetic variation and conversion}\label{sec3}

Observations (e.g., \citealt{1993MNRAS.265.1003L,2015MNRAS.446..857L}) show that the evolution of pulsar magnetic fields is slow and secular during stable spindown stages.
Therefore, the induction equation in a short duration can be reduced to
\begin{eqnarray}\label{W1}
\frac{\partial \mathbf{B}_{0}}{\partial t}=\nabla\times (\mathbf{v}\times \mathbf{B}_{0})=0,
\end{eqnarray}
where $\mathbf{B}_{0}$ is the magnetic field strength before its sudden variation,
$\mathbf{v}$ is the velocity of the charge particles as they traverse the magnetic field, and $t$ is the time.
Equation (\ref{W1}) is equivalent to the consensus that the conductivity of a pulsar, $\sigma$, is very large.
Therefore, the electric field should satisfy $\mathbf{E}=-\mathbf{v}\times\mathbf{B}_{0}$  \citep{1975clel.book.....J}, and one of Maxwell's equations
\begin{eqnarray}\label{W8}
\mathbf{j}=\frac{1}{4\pi}\left ( \nabla\times\mathbf{B}_{0}-\frac{\partial \mathbf{E}}{\partial t} \right ),
\end{eqnarray}
can be rewritten as
\begin{eqnarray}\label{W9}
\mathbf{j}=\frac{1}{4\pi}\left ( \nabla\times\mathbf{B}_{0}+\frac{\partial \mathbf{(\mathbf{v}\times \mathbf{B}_{0}})}{\partial t} \right ).
\end{eqnarray}
According to equation (\ref{W1}) and $\partial \mathbf{v}/\partial t=0$ due to the sluggish spindown in a short duration, we also have
\begin{eqnarray}\label{W99}
\frac{\partial \mathbf{E}}{\partial t}=-\frac{\partial (\mathbf{v}\times \mathbf{B}_{0})}{\partial t}=0.
\end{eqnarray}
This means that induced current vanishes in the stable spindown stage and only conduction current is left.
Substituting equation (\ref{W99}) into equation (\ref{W9}) gives
\begin{eqnarray}\label{W10}
\mathbf{j}=n\mathbf{v}=\frac{1}{4\pi}\left ( \nabla\times\mathbf{B}_{0}\right ),
\end{eqnarray}
where $n$ is the charge density of either positive or negative charges within the electroneutral plasma.
Therefore, once there is a perturbation $\mathbf{v}\rightarrow \mathbf{v}+\delta\mathbf{v}$,
the magnetic field should change as $\mathbf{B}_{0}\rightarrow \mathbf{B}_{0}+\Delta\mathbf{B}$ to maintain equation (\ref{W10}).
So, we have
\begin{eqnarray}\label{W10'}
n(\mathbf{v}+\delta\mathbf{v})=\frac{1}{4\pi}\left[ \nabla\times(\mathbf{B}_{0}+\Delta\mathbf{B})\right],
\end{eqnarray}
that is
\begin{eqnarray}\label{W10''}
n\delta\mathbf{v}=\frac{1}{4\pi}\nabla\times\Delta\mathbf{B}.
\end{eqnarray}

In above, we have shown that the magnetic field should be altered with the variation of the velocity field.
As discussed in Section \ref{sec2}, we emphasize again that the variation of the velocity field may trigger the release of the magnetic free energy in Vela pulsar (and even the stress energy; \citealt{1991ApJ...382..587R}),
that is the toroidal component may be transformed into the poloidal componet.
This problem is complicated and strong assumptions may be required (see, e.g., \citealt{Akgun} for other discussions).
Our primary focus lies in examining the trend of the magnetic evolution,
particularly during the brief period following the disturbance, wherein the magnetic variation has not yet influenced the underlying variations of the magnetic field itself.
Therefore, we simply consider the following simplified case.
(i) Before the perturbation of the velocity field, the magnetic field is static (i.e., equation (\ref{W1}) is satisfied).
(ii) After the perturbation ($\mathbf{v}\rightarrow \mathbf{v}+\delta\mathbf{v}$), the magnetic variation only evolves in response to the change in the velocity field, that is $\delta\mathbf{v}$ has nothing to do with the original magnetic field.
According to the simplified scenario,
the altering magnetic field, $\mathbf{B}$, determined by
\begin{eqnarray}\label{W222'}
\frac{\partial \mathbf{B}}{\partial t}=\nabla\times [(\mathbf{v}+\delta \mathbf{v})\times \mathbf{B}],
\end{eqnarray}
can be simplified to
\begin{eqnarray}\label{W222}
\frac{\partial\mathbf{B}}{\partial t}=\nabla\times (\delta \mathbf{v}\times \mathbf{B}).
\end{eqnarray}
Since the perturbation, $\delta\mathbf{v}$, is resulted by the transfer of angular momentum,
there should be $\delta\mathbf{v}\propto \delta\mathbf{\Omega}\times\mathbf{r}$,
where $\delta \mathbf{\Omega}=(\delta\Omega \cos\theta, \delta\Omega \sin\theta, 0)$ is set as a constant perturbation of the pulsar spin, $\mathbf{\Omega}$,
in the spherical coordinate frame $(r, \theta, \varphi)$ with $r=|\mathbf{r}|=0$ being the centre of the pulsar.
Consequently, equation (\ref{W222}) can be decomposed into
\begin{eqnarray}\label{W23}
-\frac{\partial B_{\rm r}}{\partial t}\propto \nabla_{\varphi}( B_{\rm r}\delta \Omega\; r\sin \theta ),
\end{eqnarray}
\begin{eqnarray}\label{W24}
-\frac{\partial  B_{\theta}}{\partial t}\propto \nabla_{\varphi}( B_{\rm \theta}\delta \Omega\; r\sin \theta ),
\end{eqnarray}
and
\begin{eqnarray}\label{W25}
\frac{\partial  B_{\varphi}}{\partial {t}}\propto [\nabla_{\rm r}( B_{\rm r}\delta \Omega\; r\sin \theta )+\nabla_{\theta}( B_{\rm \theta}\delta \Omega\; r\sin \theta ) ],
\end{eqnarray}
where the subscripts denote the components of vectors (operators), such as $\nabla_{\rm r}=\frac{\partial}{\partial r}$, $\nabla_{\theta}=\frac{1}{r}\frac{\partial}{\partial \theta}$ and $\nabla_{\varphi}=\frac{1}{r\sin\theta}\frac{\partial}{\partial \varphi}$.
According to equation (\ref{W23}), the variation part of $B_{\rm r}$ can be formalized as
\begin{eqnarray}\label{W27}
 \Delta B_{\rm r}= B_{1}e^{\pm \varphi}e^{\mp\delta \Omega t},
\end{eqnarray}
where $ B_{1}$ is a certain function of $r$ and $\theta$.
Similary, according to equation (\ref{W24}), the variation part of $B_{\rm \theta}$ is
\begin{eqnarray}\label{W28}
\Delta B_{\theta}= B_{2}e^{\pm\varphi}e^{\mp\delta \Omega t},
\end{eqnarray}
where $B_{2}$ is also a certain function of $r$ and $\theta$.
Substituting equations (\ref{W27}) and (\ref{W28}) into equation (\ref{W25}) gives
\begin{eqnarray}\label{W33}
\Delta B_{\varphi}=\mp \left[(\nabla_{\rm r}B_{\rm r})r\sin\theta+  B_{\rm r}\sin\theta
+(\nabla_{\theta}B_{\theta})r\sin\theta+ B_{\theta}\cos\theta \right],
\end{eqnarray}
where $\Delta B_{\varphi}$ is the variation part of $B_{\rm \varphi}$.
As shown by equation (\ref{W27}) and (\ref{W28}), it is indeed that the liberation of the magnetic free energy accumulated in the toroidal field has the potential to augment the poloidal field,
that is $\Delta B_{\rm r}\propto e^{\delta\Omega t}$ and $\Delta B_{\rm \theta}\propto e^{\delta\Omega t}$.

Certainly, the poloidal component cannot be enhanced indefinitely, and equation (\ref{W27}) and (\ref{W28}) just consider a simplified case.
An evolutive system implies that it stays at a state of excitation.
The end of the evolution should be the attainment of the minimum free energy state within the system.
A potential criterion for deciding the evolutionary trajectory of magnetic variation could involve assessing whether the enhancement of the magnetic field,
as described by equation (\ref{W27}) and (\ref{W28}), is capable of achieving the magnetic variation determined by equation (\ref{W10''}) during several dynamic time scales
of the pulsar system as indicated by the observation of \cite{2018Natur.556..219P} (another critical condition may be straining the pulsar crust beyond its elastic; \citealt{1991ApJ...382..587R}).
If the pulsar system can restrain the enhancement of the poloidal component prior to the perturbation escalating into an instability,
it should revert to its previous magnetic state. Conversely,
if the restraint is unsuccessful, the system should turn into a distinct magnetic state characterized by a reduced free energy (under this situation, the final magnetic variation, $|\Delta\mathbf{B}|$, may be much larger than the variation determined by equation (\ref{W10''})).

\section{A toy model}
\begin{center}
\begin{figure*}
\centering
\includegraphics[scale=.4]{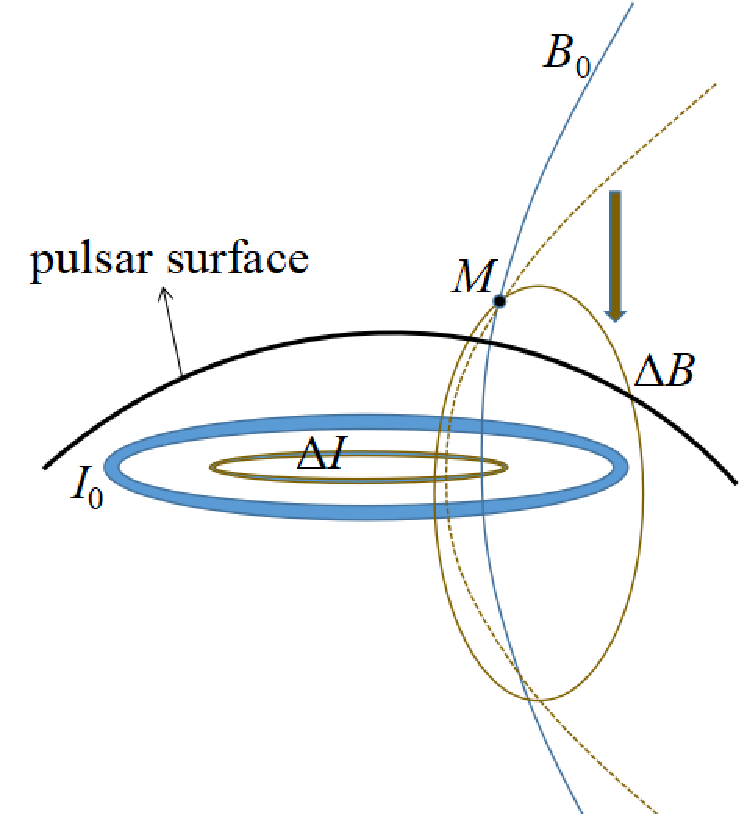}
\includegraphics[scale=.44]{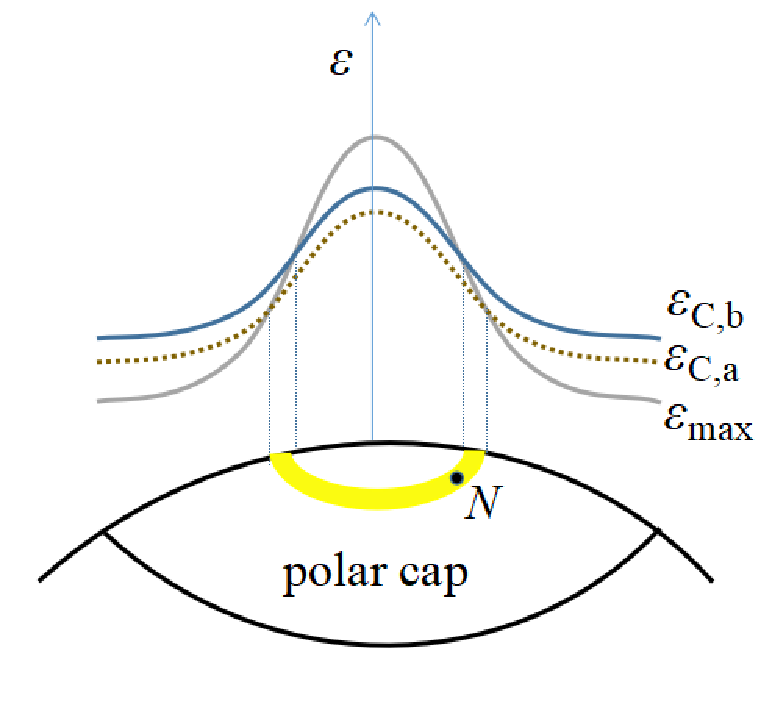}
\caption{A schematic diagram for the toy model.
\underline{Left panel}: an equivalent treatment of the generation of the pulsar magnetic field.
The blue annulus is the electric current, $I_{0}$, which generates the original magnetic field, $\mathbf{B}_{0}$. The blue solid line is the magnetic field line of $\mathbf{B}_{0}$.
The orange annulus is the perturbed electric current, $\Delta I$, which generates the varying magnetic field, $\Delta\mathbf{B}$. With the shrink of the perturbed electric current, the
magnetic field lines of $\Delta\mathbf{B}$ changes from the orange dashed line to the orange solid line.
\underline{Right panel}: the effect of the new generated magnetic field line (the dashed orange line in the left panel).
The gray solid line is the max voltage, $\varepsilon_{\rm max}$, can be supplied in the gap.
The blue solid line is the critical voltage, $\varepsilon_{\rm C,b}$, required to trigger sparks before the emergence of the magnetic variation.
The orange dotted line is the critical voltage, $\varepsilon_{\rm C,a}$, required to trigger sparks after
the emergence of the magnetic variation (corresponding to the the dashed orange line in the left panel).
The yellow wide arc is constituted of the foot points of the magnetic field lines in the increased spark region (Point $N$ is the foot point of the magnetic field line passing through point $M$).
}
\label{f2}
\end{figure*}
\end{center}

\begin{center}
\begin{figure*}
\centering
\includegraphics[scale=.45]{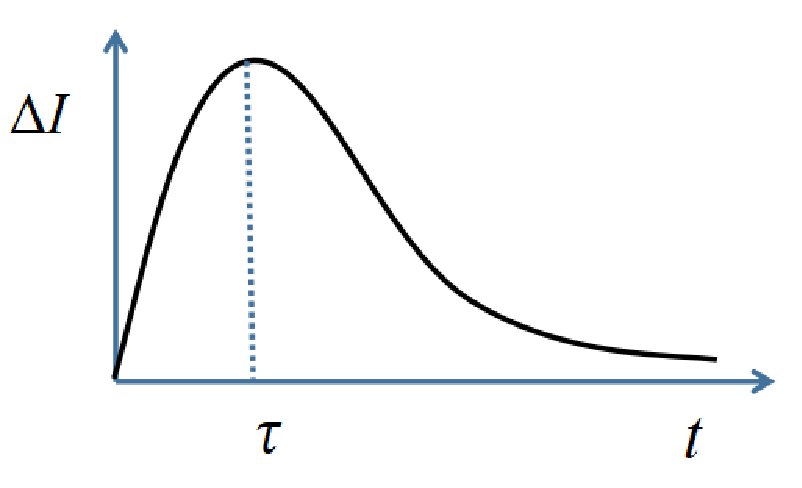}
\caption{A schematic diagram for the evolution of $\Delta I$.
}
\label{f3}
\end{figure*}
\end{center}

Based on the possibility discussed above, in this section, we construct a toy model (see Figure \ref{f2}) to elucidate the pulse broadening and pulse missing.

As the holistic magnetic structure of a pulsar is consisted of the close-field-line region and the open-field-line region, we roughly consider that the stable magnetic field of Vela pulsar is maintained by an effective ring current, $I_{0}$.
When there is a release of the free energy in the pulsar that $\mathbf{v}\rightarrow \mathbf{v}+\delta\mathbf{v}$, as well as $\mathbf{B}_{0}\rightarrow \mathbf{B}_{0}+\Delta\mathbf{B}$, we also consider that the magnetic variation is resulted by an effective perturbed electric current, $\Delta I$. Accordingly, this perturbed electric current
should be supplied by an effective voltage, $\Delta U$. Given the Lenz's law, the perturbation can be described as a resistor-inductor circuit (${R}L$ circuit), that
\begin{eqnarray}\label{eq20}
\Delta U(t)=\Delta I(t) {R}+L\frac{d\Delta I(t)}{dt}.
\end{eqnarray}

Since the rapid pulse variations observed by \cite{2018Natur.556..219P} only last for $5$ rotation periods,
the perturbed voltage should decay rapidly.
We assume that $\Delta U(t)=\Delta U_{0} e^{-\alpha t}$ with $\Delta U_{0}=\Delta U(t=0)$ and $\alpha>0$.
Then the solution of equation (\ref{eq20}) is
\begin{eqnarray}
\Delta I(t)=\frac{\Delta U_{0}}{L\alpha -R}e^{-\frac{R}{L} t}\left [ 1-e^{-(\alpha -\frac{R}{L})t}\right].
\end{eqnarray}
If the perturbed current works, its peak value, $\Delta I_{\rm max}$, should satisfy
\begin{eqnarray}
\Delta I_{\rm max}=\frac{\Delta U_{0}}{L\alpha -R}e^{-\frac{R}{L} \tau}\left [ 1-e^{-(\alpha -\frac{R}{L})\tau}\right]>I_{0},
\end{eqnarray}
where $\tau=\frac{\ln(L\alpha /R -1)}{2\alpha-R/L}$ is the peak time.
The evolution of $\Delta I$ is shown in Figure \ref{f3} with $L\alpha>R$.
According to the popular model of glitches \citep{1975Natur.256...25A,1981ApJ...248.1099A,1984ApJ...276..325A}, the initial perturbation should occur near the pulsar inner crust. So, the perturbation will transmit from the pulsar exterior to the interior. Therefore, the radius of the effective perturbed current, $\Delta I$, should shrink with time. At a certain time the radius of $\Delta I$ can be smaller than that of $I_{0}$.
So, a more curved magnetic component (the dashed orange line in the left panel of Figure \ref{f2}) would be superposed onto the original field.
According to equation (\ref{o1}), the require spark voltage is decreased, for example, from $\varepsilon_{\rm C,b}$ to $\varepsilon_{\rm C,a}$.
Following the discussion shown in Section \ref{sec2}, the Gaussian-like pulse profile implies a Gaussian-like gap, for example, the gap voltage, $\varepsilon_{\rm max}$, shown in the right panel of Figure \ref{f2}.
The decrease of the critical spark voltage will lead to part of the region where $\varepsilon_{\rm C,b}>\varepsilon_{\rm max}$ turning into the region with $\varepsilon_{\rm C,a}<\varepsilon_{\rm max}$.
Therefore, the spark region is broadened, as well as the pulse profile.
If the perturbed current continues to shrink and can still maintain a large value ($\Delta I > I_{0}$),
an extreme case may emerge that the original open field line through point $M$ gradually changes into a close field line. Under this extreme case, part of the open-field-line region will turn into a new close-field-line region. If the line of sight passes over point $M$ at this time, the pulse will disappear, since the gap can not develop in the close-field-line region. Only the perturbed current decreases to the state that $\Delta I < I_{0}$ and the magnetic field line passing through point $M$ converts back into the open field line, can the pulse be seen again.

The above toy model is a simplified one, the related parameters are hardly to be determined.
However, there are two basic requirements should be satisfied. (a) The peak time of the perturbed current should be $\tau \sim 2P$. (b) The magnetic variation, $\Delta\mathbf{B}$, should be sufficiently strong. The second requirement seems to be an overclaim. Several reasons may mitigate this difficulty. (I) The toroidal component of the magnetic field may be much stronger than the poloidal component. The transient variation is strong for the poloidal component, but relatively weak for the toroidal component. (II) The toy model is an idealized model, many details are ignored. For example, in some places, the original field may be more suited to be equivalent to the field generated by the current in a solenoid. The original field can be sparse in some places, so that a small $\Delta\mathbf{B}$ can still work. (III) The perturbation may be localized in a small area. In this area, the magnetic variation, $\Delta\mathbf{B}$, may be stronger than the original field since a temporary strong change in the local magnetic field may not necessarily result in a comprehensive transformation of the entire field.
(IV) It could be merely the direction of the magnetic field that undergoes a change, while the magnitude remains approximately constant, i.e., $|\mathbf{B}_{0}|\approx |\mathbf{B}_{0}+\Delta\mathbf{B}|$. The necessary change into the magnetic energy could be small.

\section{Summary and discussion}
In this manuscript, according to the alterations in the pulse profile and polarization of Vela pulsar associated with the glitch that occurred on December 12, 2016,
we ratiocinate how the magnetic field of Vela pulsar may undergo alterations. The variations in the magnetic field which is induced by changes in the velocity field
serves as the underlying cause for the alterations in the pulse shape.
Specifically, the liberation of the free energy stored within the toroidal component of the magnetic field induced by changes in the velocity field leads to the emergence of new poloidal component.
In such a scenario, the pulsar system is considered to be in a magnetic ``excited'' state, where the newly formed poloidal component is primarily consisted of multipole components.
Therefore, the poloidal component becomes more curved, reducing the voltage required for spark generation.
Correspondingly, in the polar cap region, the spark area increases and the pulse profile is broadened.
Furthermore, if the newborn poloidal component further develops and the open magnetic field lines change into close field lines, part of open-field-line region will be transformed into a new close-field-line region. The gap cannot grow in this newly formed close-field-line region, so that there will be the observed pulse missing when the line of sight sweep over this region.
To visualising the proposed physical picture, we also construct a toy model: the magnetic variation is equivalently treated as the perturbation of an annular electric current, $\Delta I$.

In above picture, we only discuss the effect imposing on pulsed radio emissions of the short-term changes in pulsar magnetic fields. Two extra possible consequences of magnetic variations are worthy of paying attentions.
\begin{itemize}
\item According to \cite{1969ApJ...157..869G}, for the charge particles in the steady magnetic field, $\mathbf{B}_{0}$, of a pulsar, the force inflicted by the electric field, $\mathbf{E}_{\rm p}$, is balanced by the Lorentz force, that is
\begin{eqnarray}\label{o2}
\mathbf{E}_{\rm p}+\frac{(\mathbf{\Omega} \times \mathbf{r}\times \mathbf{B}_{0})}{c}=0,
\end{eqnarray}
where $c$ is the speed of light.
When the perturbation $\mathbf{B}_{0}\rightarrow \mathbf{B}_{0}+\Delta \mathbf{B}$ happens, equation (\ref{o2}) is briefly changed to
\begin{eqnarray}\label{eq2}
\mathbf{E}_{\rm p}+\frac{[\mathbf{\Omega} \times \mathbf{r}\times (\mathbf{B}_{0}+\Delta\mathbf{B})]}{c}\neq 0.
\end{eqnarray}

Therefore, there will be an induced electric field with a potential which can be roughly estimated as
\begin{eqnarray}\label{w35}
\Delta V&\sim&\frac{2\pi P |\Delta \mathbf{B}| r_{\ast}^{2}}{c}
=2.1\times 10^{13} P_{10}^{-1}\eta_{-2}B_{0,14}r_{\ast,6}^{2}{\rm V},
\end{eqnarray}
where $P_{6}=P/10\rm {s}$, $r_{\ast}$ is the star radius and $r_{\ast,6}=r_{\ast}/10^{6}{\rm cm}$, and ${|\Delta \mathbf{B}|}=\eta|\mathbf{B}_{0}|=\eta B_{0}$ with $\eta_{-2}=\eta/10^{-2}$ and $B_{0,14}=B_{0}/10^{14}{\rm G}$.\footnote{Note that, in equation (\ref{w35}), $|\Delta \mathbf{B}|$ may be induced by other reasons than just the variation of the velocity field, for example, the XRB itself.} Therefore, provided that the effective magnetic variation of a magnetar is merely $1\%$ of the primary magnetic field, a more energetic electric field will manifest on the magnetar (e.g., the magnetar SGR 1935+2154), surpassing that of the inner gap ($\sim 10^{9}\rm V$ ), until the induced electric field is screened by the rearrangement of electric charges.

$\;\;\;\;$We note that certain observations reveal a correlation between X-ray bursts (XRBs; \citealt{Mereghetti,Li,Ridnaia,Tavani}), glitches \citep{2024Natur.626..500H}, antiglitches (spin-down glitches; \citealt{Younes,GMY}) and fast radio bursts (FRBs) \citep{CHIME,Bochenek}.
As XRBs, glitches and antiglitches are potentially linked to alterations in the magnetic fields of magnetars, FRBs may be also the consequences of magnetic variations. The statistic analyses also indicate the connection between magnetar short bursts, as well as magnetic variations, and FRBs \citep{2019ApJ...879....4W,2020ApJ...891...82W}. A key problem for understanding FRBs is that how the charge particles which emit coherent radio emissions are accelerated.
As of now, no conclusive observation showing that a repeating FRB possesses a periodicity close to rotational periods of pulsars \citep{2018ApJ...866..149Z,2021Natur.598..267L}. This indicates that the transient electric fields accountable for FRBs differ from the electric fields within inner gaps. Given the above connections, the electric fields given by equation (\ref{w35}) may be the ones required for the occurrences of FRBs.
\end{itemize}

\begin{itemize}
\item
Conversely, if the magnetic variation is long-term (e.g., the state shown by the dashed orange line in Figure \ref{f2}),
the reduction of the critical voltage required for spark generation can be maintained for a long time. This may lead to the reactivation of certain radio-quiet pulsars  (\citealt{2006Natur.442..892C,2007ApJ...666L..93C,2020ATel13699....1Z,2023SciA....9F6198Z}; one can also refer to \citealt{Dong,Maan,Huang}).
\end{itemize}

\section*{Acknowledgements}
I appreciate the reviewer's valuable comments.
I thank Dr. Wei-Hua Wang for introducing glitches for me very much.
This work is supported by a research start-up grant from Tongling University (R23052).


\appendix
\section{}
According to magnetostatics, multipole components should emerge on pulsars. A slow rotating pulsar should closely resemble a sphere. The presence of a pure magnetic dipole field on the pulsar implies that the sphere should be uniformly magnetized. However, for a pulsar possessing a hybrid magnetic field comprising both poloidal and toroidal components, achieving uniform magnetization becomes challenging. Magnetic multipole components should also be present on the pulsar. The stable magnetic configuration on a pulsar necessitates the coexistence of both dipole component and multipole components (``magnetic ground state''). A pulsar with an altering magnetic field can be considered to be transitioning from a magnetic excited state to its magnetic ground state. Consequently, during the transformation of the toroidal component into a new poloidal component, the generation of new multipole fields should be inevitable since the magnetic ground state involves multipole fields.

\end{document}